\documentclass[aps,prl,twocolumn,preprintnumbers,nofootinbib,superscriptaddress,amsmath]{revtex4-2}
\usepackage[english]{babel}
\usepackage[pdftex, pdftitle={Article}, pdfauthor={Author}]{hyperref} 
\usepackage{graphicx}
\usepackage{dcolumn}
\usepackage{bm}
\usepackage{float}
\usepackage{orcidlink}
\usepackage{subcaption}
\usepackage{booktabs,siunitx}

\begin{document}


\title{Skipping the rungs! Calibrating distance indicators through their clustering with galaxies}

\author{Madeline L. Cross-Parkin \orcidlink{0009-0002-9780-1869}}
\altaffiliation{These authors contributed equally to this work.}

 \affiliation{School of Mathematics and Physics, The University of Queensland,
 Brisbane, QLD 4072, Australia}
 \affiliation{OzGrav: The ARC Centre of Excellence for Gravitational Wave Discovery }
\author{Leonardo Giani  \orcidlink{0000-0001-6778-1030}}
\altaffiliation{These authors contributed equally to this work.}
\affiliation{Swinburne University of Technology, Hawthorn VIC 3122, Australia}
\affiliation{OzGrav: The ARC Centre of Excellence for Gravitational Wave Discovery }


 \author{Cullan Howlett\orcidlink{0000-0002-1081-9410}}
\affiliation{School of Mathematics and Physics, The University of Queensland,
 Brisbane, QLD 4072, Australia}
 \affiliation{OzGrav: The ARC Centre of Excellence for Gravitational Wave Discovery }

\author{Tamara M.~Davis \orcidlink{0000-0002-4213-8783}}
\affiliation{School of Mathematics and Physics, The University of Queensland,
 Brisbane, QLD 4072, Australia}
 \affiliation{OzGrav: The ARC Centre of Excellence for Gravitational Wave Discovery }





\date{\today}
\begin{abstract} 
We show that angular cross-correlations between distance indicators and galaxy redshift catalogues, when interpreted within an assumed cosmological model, can constrain potential biases in distance measurements induced by calibration systematics. As a test case, we consider a simple scenario in which a constant calibration offset $\Delta M$ shifts all observed distances by a multiplicative factor, and we produce Fisher forecasts for the constraining power on $\Delta M$ and $H_0$ from existing and upcoming surveys. In our most optimistic scenario, based on the expected number of SN~Ia observed by LSST and the DESI final data release, we find $\sigma_{\Delta M} \approx 0.05$ and $\sigma_{H_0} \approx 1.81~\mathrm{km~s^{-1}~Mpc^{-1}}$. This has important implications for the Hubble tension, since explaining the discrepancy purely as a calibration systematic in low-$z$ measurements would require $\Delta M \approx 0.13$.
\end{abstract}
\maketitle

\section{Introduction}
A famous quote by Sandage \citep{1970Sandage} summarised the state of observational cosmology in the 1970s as “a search for two numbers”: the present-day expansion rate of the Universe, the Hubble constant $H_0$, and the deceleration parameter $q_0$. Half a century later, this search has culminated in some of the most serious challenges yet faced by the standard cosmological model. Chief among them is the Hubble tension \citep{Verde:2019ivm,CosmoVerseNetwork:2025alb,Abdalla:2022yfr}, a persistent incompatibility between late- and early-Universe determinations of $H_0$. Its significance is often emphasised by quoting the striking $\sim5 \sigma$ discrepancy between the local Cepheid-calibrated type Ia supernovae (SN~Ia) measurements of the SH0ES collaboration \citep{Riess:2021jrx} and the value inferred from Cosmic Microwave Background (CMB) observations by the Planck collaboration \citep{Planck:2018vyg}. Even if less significant, the tension persists between multiple other cosmological probes, hinting that its root is unlikely to rely solely on the physical nature (or calibration) of any specific tracer.
In the realm of its possible explanations, the two most popular lines of thought are either unknown systematics in the measurements or new physics beyond the vanilla $\Lambda$CDM \cite{Hu:2023jqc,Vagnozzi:2019ezj,Vagnozzi:2023nrq,DiValentino:2021izs}.
\textit{Systematics or new physics?} has therefore emerged as the most compelling question facing the community. 

In this Letter, we introduce an innovative methodology to address the first of these possibilities. Under the assumption that no new physics is at play, we construct an observable capable of exposing potential systematic biases of distance indicator measurements. We specialise to the following scenario: a generic sample of standard candles, consisting of distance measurements $d_i^{\mathrm{obs}}$ and redshifts $z_i$, has been incorrectly anchored via some primary distance indicator, resulting in inferred distances systematically biased by a constant multiplicative factor. For reference, consider a set of SN~Ia distance moduli measurements $\mu_i$ converted into luminosity distances $d_{L,i}$ assuming a ``wrong'' fiducial absolute magnitude $M_{\rm fid}$
\begin{equation}
    \mu_i = m_i - M_{\rm{fid}} = 5\log_{10}{d_{L,i}^{\rm{obs}}} +25 \;, \quad M_{\rm{true}}=M_{\rm{fid}} + \Delta M\;.
    \end{equation}
Since $d_L\propto H_0^{-1}$, distance measurements alone are strongly degenerate in the parameters $H_0, \Delta M$.

Although simplistic, such scenario is particularly relevant. Indeed, it has been shown that the Hubble tension can be described as a ``magnitude tension'' \citep{Camarena:2023rsd}, and if the SN~Ia are anchored to high redshift calibrators (a technique usually referred to as \textit{inverse distance ladder}, see for example Refs.~\cite{DES:2024ywx,Camarena:2019rmj}) rather than Cepheids, the discrepancy with Planck disappears.\footnote{This does not exclude the possibility of systematic biases in the calibration of the sound-horizon scale at high redshift. On the contrary, it further motivates consistency tests between low- and high-redshift calibration, such as the one presented here.} We demonstrate that the angular distribution of the SN~Ia, unlike the standard radial distance--redshift relationship alone, contains information that breaks the degeneracy between $H_0$ and $\Delta M$, and in doing so, self-calibrates the SN~Ia. We include in the analysis an independent galaxy redshift catalogue to reduce the intrinsic Poisson uncertainties arising from discrete number counts statistics.
\begin{figure*}[t!]
\begin{center}
    \includegraphics[width=0.7\textwidth]{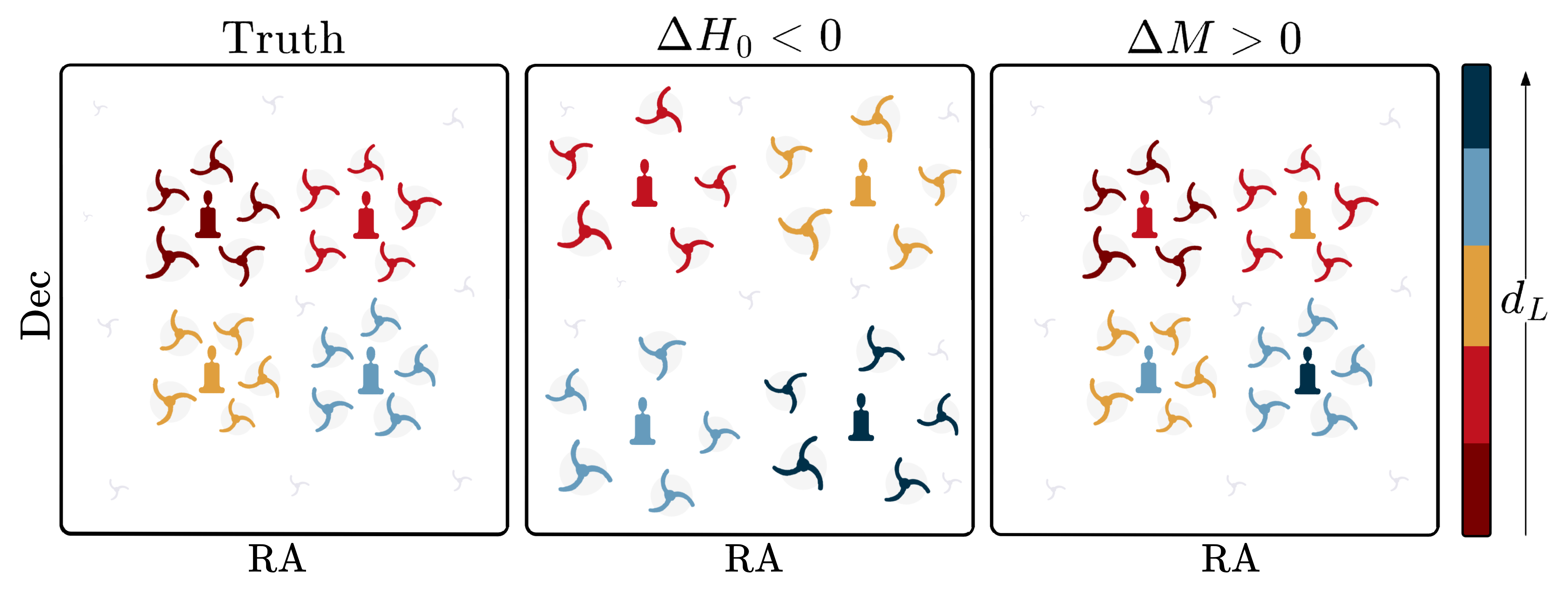}
    \caption{Pictorial representation of the AP-like distortion induced by a constant shift $\Delta M$ and a variation in the Hubble constant, $H_0$. The left panel shows the fiducial, or ``true”, distribution of galaxies and standard candles. If $H_0$ is decreased while all other cosmological parameters are held fixed, both radial and transverse comoving distances increase. This changes the apparent three-dimensional clustering pattern and dilutes the density field, as shown in the middle panel. By contrast, a positive magnitude offset, $\Delta M>0$, changes only the inferred radial distances of the standard candles; their angular positions remain fixed, and the galaxy field is unchanged.}
    \label{fig:Diagram}
\end{center}
\end{figure*}
\label{sec:Methods}
By suitably generalising previous analyses of cross-correlations in redshift and distance space (see e.g. Refs.~\cite{Oguri_2016,Ferri_2025,Mukherjee:2020hyn,Mukherjee:2018ebj,Cross-Parkin:2026wyz,Cheng:2026atn}), our results show that any systematic biases in the calibration of a set of standard candles strongly affect their clustering statistics. 

This arises because calibration systematics shift the inferred radial positions of the SN~Ia, leaving their angular positions unchanged. In contrast, a change in the expansion rate, $H_0$, affects both radial and transverse distance scales, and therefore distorts the apparent clustering pattern. Mathematically, this is analogous to an Alcock--Paczy\'nski (AP)-like distortion \cite{Alcock:1979mp,2011MNRAS.418.1725B}, as illustrated schematically in Fig.~\ref{fig:Diagram}. This distinction allows the effects of $H_0$ and calibration systematics to be partially disentangled, effectively anchoring the SN~Ia to the clustering of the matter field rather than to primary distance indicators alone.

\section{Calibration systematics and angular clustering}
Our aim is to exploit the different clustering behavior of the distance indicator field in redshift space and in observed distance space, in combination with a galaxy redshift catalog, to constrain calibration systematics. A useful statistic to study the clustering of tracers is the angular cross-correlation. In Fourier space, the cross-correlation $C_\ell^{A_iB_j}$ between tracers $A$ and $B$ in bins $i$ and $j$, respectively, can be thought of as a statistical measurement of their overlap weighted by the matter power spectrum, $P(k)$ (see Appendix~\ref{app:Cell_theory} for a brief review of angular cross-correlations).

To simplify the notation, we denote the distance-space and redshift-space distance indicator fields with the labels ${\mathrm{d}, \mathrm{s}}$, while $\mathrm{g}$ denotes the galaxy redshift catalog.
Following our SN~Ia example, consider a catalog of redshifts $z_i$, observed magnitudes $m_i$, and sky positions $(\mathrm{RA}, \mathrm{Dec})$. After standardisation, if the fiducial absolute magnitude differs from the true value according to
$M_{\mathrm{fid}}= M_{\mathrm{true}}+ \Delta M$,
for example because of unknown calibration systematics, the inferred luminosity distances are biased by a factor $\alpha_{\Delta M}$,
\begin{equation}\label{d_obsvs_true}
d_L^{\mathrm{obs}}=d_L^{\mathrm{true}}\alpha_{\Delta M} \;, \qquad \alpha_{\Delta M}=10^{\frac{\Delta M}{5}}\;.
\end{equation}

This systematic bias also affects the \textit{observed} density contrast field of the distance indicator, which can be written in terms of the \textit{true} one as
\begin{equation}\label{eq:delta}
    \delta_{d}^{\mathrm{obs}}(d^{\mathrm{obs}})= \delta_{d}^{\mathrm{true}}\left(\frac{d^{\mathrm{obs}}}{\alpha_{\Delta M}}\right)\;,
\end{equation}
see Appendix~\ref{app:delta} for the derivation. The density contrasts of the galaxy and distance indicator fields can be used to build the angular cross-correlations $C_\ell^{\rm gs}$ and $C_\ell^{\rm gd}$. The impact on $C_\ell^{\mathrm{g}_i\mathrm{d}_j}$ induced by the shift $\alpha_{\Delta M}$ is perhaps more transparent in the Limber approximation (which simplifies the line-of-sight integrals entering the angular power spectra; see Eq.~\eqref{eq:Limber_approx}):
\begin{equation} \label{eq:Cell_Limber_DM}
\begin{split}
    C_\ell^{\mathrm{g}_i \mathrm{d}_j}\approx& 
    \int_{\Delta z_{\mathrm{ov}}}dz\frac{b_{\rm{g}}(z)b_{\rm{d}}(z^*)H(z^*)}{\alpha_{\Delta M} \chi^2(z)}\\
    &\times\frac{dn_\mathrm{g}^{\rm{obs}}(z)}{dz}\frac{dn_\mathrm{d}^{\rm{obs}}(z^*)}{dz} P\left(k_{\mathrm{\ell}},z,z^*\right)\;.
\end{split}
\end{equation}
In the above expression, $dn_x^{\rm{obs}}/dz$ denotes the averaged redshift distribution of tracer $x$, and $b_{x}$ its clustering bias with respect to the underlying matter distribution. In this work we adopt a linear bias model, although this can be easily generalised to more complex parametrisations. $k_{\ell}=(\ell + 1/2)/\chi$ is the Limber scale, and the quantity $z^*=\chi^{-1}\left[\chi(z)/\alpha_{\Delta M}\right]$ is the redshift corresponding to the shifted luminosity distance. The integral has support only over the overlap region $\Delta z_{\mathrm{ov}}$ between the galaxy and shifted distance indicator bins.
We can conclude from Eq.~\eqref{eq:Cell_Limber_DM} that $\Delta M \neq 0$ affects the clustering signal in three ways: it shifts the observed number density in distance space, $dn_{\rm d}(z^\ast)/dz$; it changes the overlap between the redshift-selected and distance-selected tracers; and it modifies the Jacobian of the transformation between the shifted radial coordinate and redshift, $H(z^\ast)/\alpha_{\Delta M}$. In contrast, a change in $H_0$ would modify all the cosmology-dependent quantities appearing in the integral, for example the Limber mode $k_\ell$, whereas $k_\ell$ is independent of $\Delta M$ for fixed fiducial cosmology. In this sense, the calibration shift produces an AP-like distortion: radial and transverse information respond differently to a change in $\Delta M$.

\section{A $2\times2$ cross-correlation pipeline}
In this work, we consider 9 uniform redshift bins, although the methodology can be straightforwardly generalised to other binning choices. We then define corresponding distance bins for the distance indicators as follows. For a redshift bin $i$, with edges $z_i$ and $z_{i+1}$, containing $n$ distance indicators of type $\mathrm{s}$,
\begin{equation}
    \mathrm{s}_i = \left\{z_n|\;  z_i\leq z_n < z_{i+1} \right\}\;,  
\end{equation}
we associate a corresponding distance bin
\begin{equation}
    \mathrm{d}_i = \left\{ d^{\mathrm{obs}} (z_n) \ \forall z_n \in s_i\right\}\;.
\end{equation}
Notice that this choice implies that redshift bins do not overlap by construction, but distance bins will in general overlap because of either systematics or observational uncertainties. However, the one-to-one mapping of objects in redshift and distance ensures that overlapping distance bins $d_i$ and $d_j$ have no objects in common by construction.
We build a combined data-vector $\mathbf{D} =\left\{ {C_{\ell}^{\mathrm{g}_i\mathrm{s}_j} , C_{\ell}^{\mathrm{g}_i\mathrm{d}_j}}\right\} \; \forall\; i, j$, and define the Gaussian likelihood

\begin{equation}\label{eq:likelihood}
    \ln\mathcal{L} = -\frac{1}{2}\left[\mathbf{D}_{\mathrm{obs}} - \mathbf{D}_{\mathrm{theory}}(\theta)  \right]\Sigma^{-1}\left[\mathbf{D}_{\mathrm{obs}} - \mathbf{D}_{\mathrm{theory}}(\theta)  \right]^{\mathbf{T}}\;,
\end{equation}
where $\theta=\left\{\Delta M, H_0,\Omega_m, \lambda_m,\mu_n \right\}$ is the set of parameters entering the likelihood. We separate these parameters into those we aim to constrain, $\left\{H_0, \Delta M\right\}$; those we fix to their fiducial values, $\lambda_m =\left\{A_s, N_{\mathrm{eff}}, n_s, \Omega_b, \Omega_r\right\}$; and those we marginalise over, $\mu_n =\left\{\Omega_m, b_{\mathrm{g}}, b_{\mathrm{d}}\right\}$.
We assume Gaussian fields and define an analytical covariance with matrix elements
\begin{equation}\label{eq:covariance}
\begin{split}
    &\Sigma(C_{\ell}^{ij}, C_{\ell'}^{lm}) =\frac{\delta_{\ell \ell'}}{f_{\rm sky}\left(2\ell + 1\right)} \\
    \times&\left[\left(C^{i l}_{\ell} + \mathcal{N}^{i l}_{\ell}\right)\left(C^{j m}_{\ell'}+\mathcal{N}^{j m}_{\ell'}\right) + \left(C^{im}_{\ell}C^{jl}_{\ell'}\right)^2\right]\;,
\end{split}
\end{equation}
where the indices $i,j,l,m$ run over ${\mathrm{g},\mathrm{s},\mathrm{d}}$; $f_{\rm sky}$ is the sky fraction over which the tracers overlap, and $\mathcal{N}^{ij}_\ell$ are the shot-noise spectra of the fields (see Appendix~\ref{app:Covariance} for the details of the derivation). Notice that our binning choice implies $\mathcal{N}^{\mathrm{d}_i\mathrm{s}_i}\neq 0$, since the corresponding redshift and distance bins contain the same objects by construction.

\section{Fisher Forecasts}\label{sec:Fisher}
\begin{figure}[t]
    \centering
        \includegraphics[width=\linewidth]{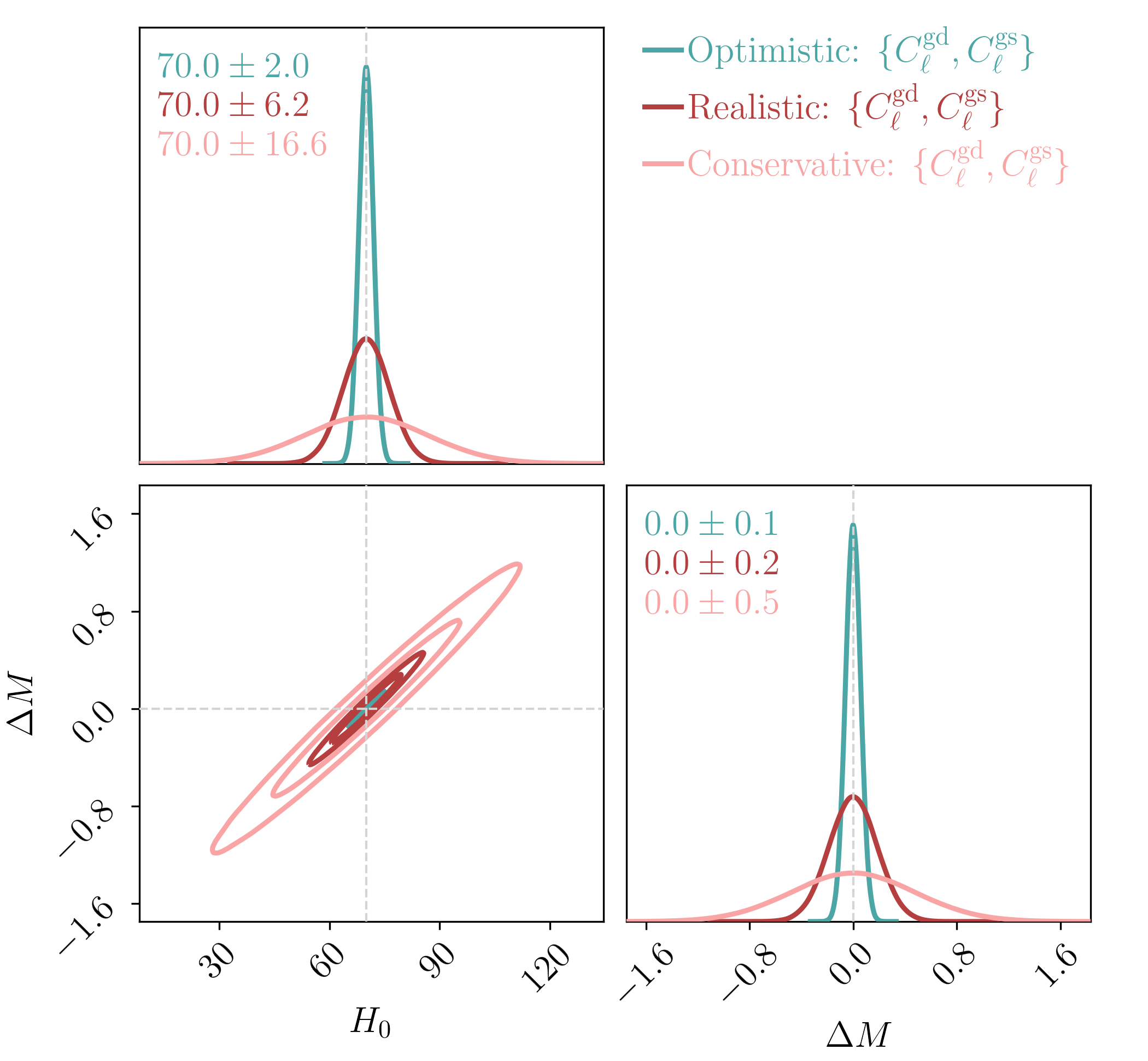}
        \caption{Fisher forecast on $H_0$ and $\Delta M$ from the likelihood in Eq.~\eqref{eq:likelihood} for the conservative, realistic, and optimistic configurations.}
        \label{fig:forecast_joint}
\end{figure}
We compute forecasts for the likelihood and covariance in Eqs.~\eqref{eq:likelihood},\eqref{eq:covariance} using \texttt{CAMB} \cite{2011ascl.soft02026L}, and assess the constraining power of our pipeline using the Fisher information matrix \citep{Fisher:1922saa} (see Eq.~\eqref{eq:fisher}). We focus on low-redshift galaxies and SN~Ia catalogs $z\leq 0.3$ specified by the number of tracers $N_{\mathrm{SN}},N_{\mathrm{gal}}$ and the uncertainties on redshift measurements $\sigma_z$ and distance measurements $\sigma_d = \kappa d $ (with  $\kappa$ constant parameterising the scaling of the errors with the distances). The Fisher matrix includes $H_0$, $\Delta M$, $\Omega_m$, and the constant
tracer biases $b_\mathrm{g}$ and $b_\mathrm{d}$ as free parameters. All
remaining cosmological parameters are fixed to their fiducial values.\footnote{
$A_s = 2.1 \times 10^{-9}$, $\Omega_b = 0.0478$, $n_s = 0.95$,
$N_{\mathrm{eff}} = 3.044$, and $T_{\mathrm{CMB}} = 2.71~\mathrm{K}$.}
The final forecasts on the parameter subspace $\left\{H_0, \Delta M\right\}$ are obtained by considering only the corresponding blocks of the inverse Fisher matrix, $F^{-1}$. We adopt conservative scale cuts with $\ell_{\min}=10$, while the maximum multipole is defined by requiring $k\leq k_{\max}=0.1\,h\,{\rm Mpc}^{-1}$. This avoids contamination from non-linear modes and ensures the consistency of the Limber computation. As a validation check, we have repeated the calculation without the Limber approximation in selected cases and verified that it remains accurate over the adopted scale cuts. We also assume spectroscopic redshifts throughout, such that the redshift uncertainty is negligible compared to the bin widths used in the analysis. Our Fisher forecasts are shown in Fig.~\ref{fig:forecast_joint} and summarised in Table~\ref{tab:fishersummary} for the following scenarios:
\begin{itemize}
    \item \textbf{Conservative:} $N_{\mathrm{gal}}=10^6$, $N_{\mathrm{SN}}=10^3$ and $\kappa = 0.1$. This scenario aims to represent the constraining power achievable with existing galaxy catalogs from DESI \citep{DESI_Collaboration_2022}, and a supernovae catalog such as ZTF \citep{bellm2014zwickytransientfacility}, assuming a sky overlap $f_{\mathrm{sky}}=0.3$.

    \item \textbf{Realistic:} $N_{\mathrm{gal}}=10^6$, $N_{\mathrm{SN}}=10^4$ and $\kappa = 0.05$. This corresponds to the constraining power achievable by combining DESI with the expected number of LSST SN~Ia \cite{Howlett:2017asw,LSST:2008ijt} detected during the first few years of observations, within their overlapping sky area of approximately $f_{\mathrm{sky}}=0.1$.

    \item \textbf{Optimistic:} $N_{\mathrm{gal}}=10^7$, $N_{\mathrm{SN}}=10^5$  and $\kappa = 0.05$. This scenario represents the constraining power achievable with combining DESI and 4HS \cite{2023Msngr.190...46T} galaxies with optimistic expectations from both LSST and ZTF within an almost complete sky coverage $f_{\mathrm{sky}}=0.8$.
\end{itemize}
\begin{table}[t]
    \centering
    \renewcommand{\arraystretch}{1.4}
    \setlength{\tabcolsep}{10pt}
    \begin{tabular}{lccc}
    \textbf{} 
        & \textbf{Conservative} 
        & \textbf{Realistic} 
        & \textbf{Optimistic} \\
    \midrule
    $H_0$       & $\pm16.64$  & $\pm6.23$  & $\pm1.96$ \\
    $\Delta M$  & $\pm0.47$  & $\pm0.18$  & $\pm0.06$  \\
    $\Omega_m$  & $\pm0.07$ & $\pm0.02$ & $\pm0.01$ \\
    \bottomrule
    \end{tabular}
    \caption{Marginalised $1\sigma$ constraints on cosmological parameters from the $\{C_\ell ^{\mathrm{gd}}, C_\ell^{\mathrm{gs}}\}$ analysis.}
    \label{tab:fishersummary}
\end{table}

\section{Validation on Mocks}\label{sec:Mocks}
In order to test the robustness of the Fisher analysis, we perform a simple simulation study by constructing a lognormal mock catalogue of tracers over a sky fraction $f_{\rm sky}=1/8$, assuming the same fiducial cosmology. We populate it with $N_{\rm{gal}} =\mathcal{O}(10^6)$ galaxies and $N_{\rm SN}=\mathcal{O}(10^4)$ SN~Ia ($\kappa=0.05$) below redshift $z\leq 0.3$ (i.e. similar to the realistic scenario). We measure the angular power spectra using the \texttt{NaMaster} package\footnote{\href{https://namaster.readthedocs.io/en/latest/}{https://namaster.readthedocs.io/en/latest/}.} \citep{Alonso_2019}, using 6 redshift/distance bins and restricting the measurement to the multipole range $16\leq \ell \leq 124$. We perform an MCMC exploration of the parameter subspace $\left\{H_0,\Delta M\right\}$ using the \texttt{emcee} ensemble sampler \citep{Foreman_Mackey_2013} and \texttt{ChainConsumer}\footnote{\href{https://samreay.github.io/ChainConsumer/}{https://samreay.github.io/ChainConsumer/}} \cite{Hinton2016}, fixing the remaining parameters to their fiducial values.

\begin{figure}[t!]
    \centering
    \includegraphics[width=\linewidth]{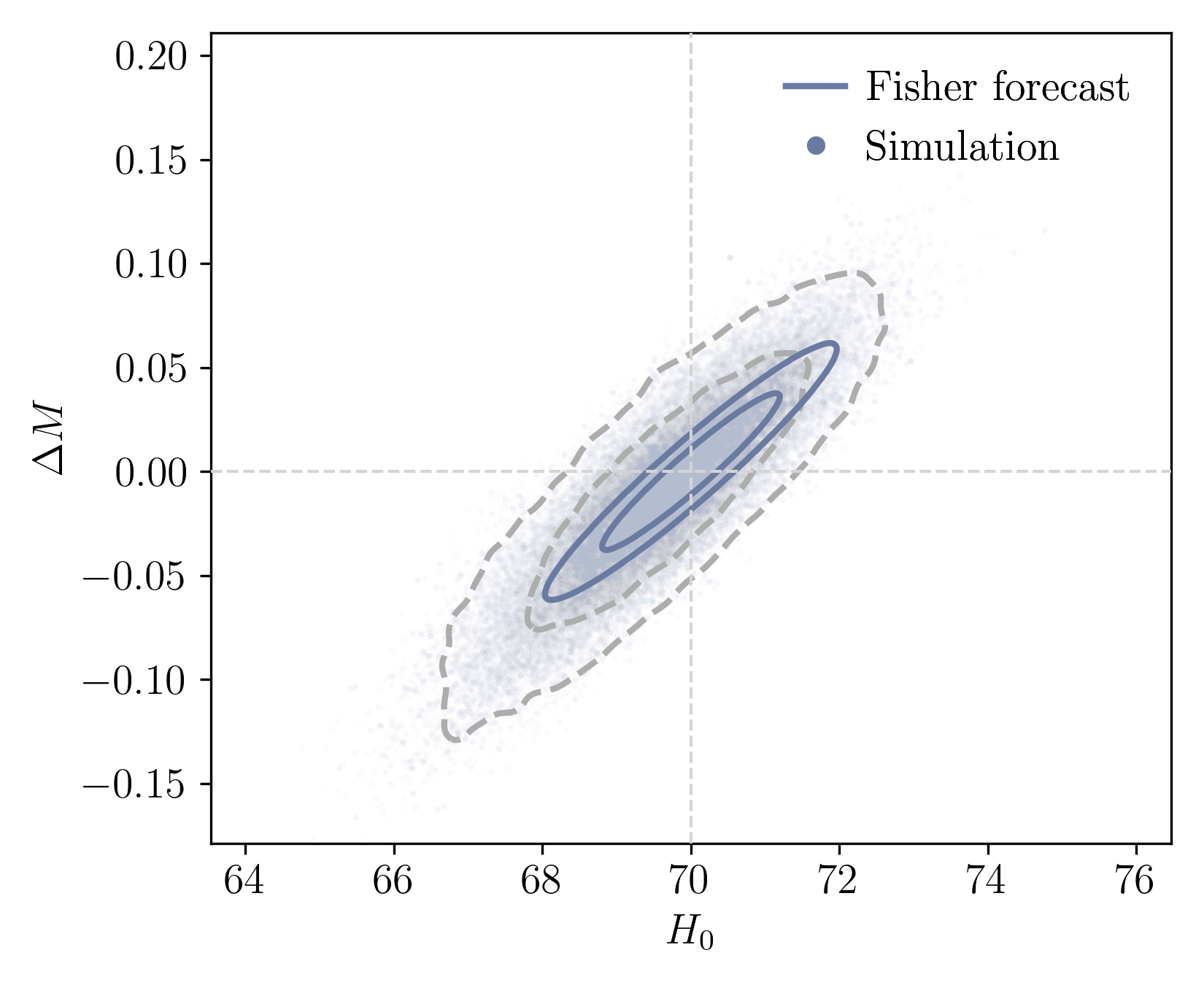}
    \caption{Comparison the Fisher forecast (solid contour) and the MCMC exploration of the simulated catalogs (scatter points and dashed contour) for the parameter space $\left\{H_0,\Delta M\right\}$ using the likelihood in Eq.~\eqref{eq:likelihood} and fixing the remaining parameters to their fiducial values.}
    \label{fig:simsvsfisher}
\end{figure}

Fig.~\ref{fig:simsvsfisher} demonstrates that the Fisher forecast and the simulation study yield consistent results. As expected, the Fisher analysis gives tighter constraints, with the simulation uncertainties being larger by a factor of approximately $1.6\,(1.8)$ in $H_0\,(\Delta M)$.

\section{Discussion and Outlook}\label{sec:Conclusions}

This Letter introduces two key novelties with respect to previous cross-correlation analyses of distance indicators and galaxy clustering: the explicit inclusion of a calibration systematic $\Delta M$ and the joint analysis of the distance indicator clustering in both redshift and distance space. As shown in Fig.~\ref{fig:forecast_joint}, the resulting methodology applied to state-of-the art SN~Ia and redshift surveys has the potential to constrain $\sigma_{\Delta M}$ with accuracy $\leq |0.1|$ in the most optimistic scenario. This could have important implications for the Hubble tension, which requires $\Delta M \approx 0.13$ to reconcile the SH0ES and the Planck values.

\begin{figure}[t]
    \centering
        \includegraphics[width=\linewidth]{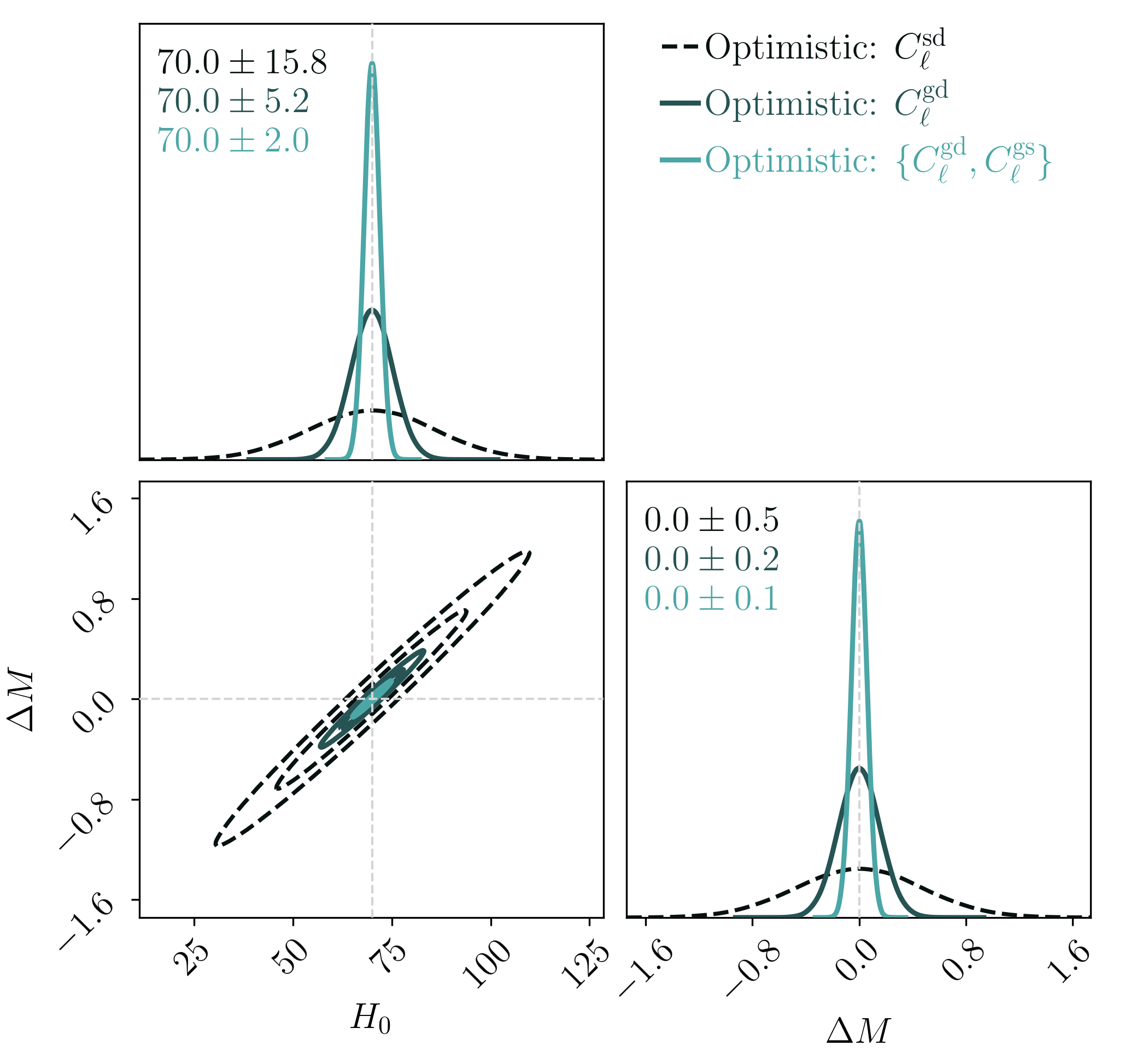}
        \caption{Marginalised constraints on $H_0$ and $\Delta M$ in the optimistic configuration, comparing the individual and joint angular power-spectrum combinations.}
        \label{fig:reduced_analisis}
\end{figure}

We assess the impact of the modelling choices introduced in our pipeline using two complementary limiting cases. In the first, we exclude the galaxy redshift catalogue and use only the clustering version of the Hubble diagram, $C_\ell^{\rm sd}$. In the second, we exclude the redshift information for the distance indicators and use only $C_\ell^{\rm gd}$. Fig.~\ref{fig:reduced_analisis} shows the forecasts for these two configurations in the \textit{optimistic} scenario. Interestingly, the analysis combining both $C_\ell^{{\rm gs}},C_\ell^{{\rm gd}}$, results in a factor of $\sim 2$ improvement on the marginalised constraints with respect to the analysis with $C_\ell^{{\rm gd}}$ only, highlighting the role of the known redshifts of the distance indicators in breaking the $H_0-\Delta M$ degeneracy. 

The clustering analysis using SN~Ia only, unsurprisingly, results in weak constraints on the parameters $\left\{H_0,\Delta M\right\}$, which we mostly attribute to the high shot-noise arising from the low number density of sources. With comparable number densities, the two approaches led to similar results, yet the analysis using $C_\ell^{\mathrm{sd}}$ will always have larger noise budget because of the cross-shot noise term $\mathcal{N}^{\mathrm{sd}}$, absent in the $C_\ell^{\mathrm{gd}}$ case.   

A word of caution concerning the importance of the parameters we kept fixed in the analysis and of the ones we marginalised over. The bias parameters are essentially unconstrained in the minimal $2\times2$ pipeline, and have almost no impact in our forecasts. In principle, information on these can be extracted by increasing the complexity of the analysis and including in our data-vector the auto-correlations of the various tracers, $C_\ell^{\mathrm{gg}},C_\ell^{\mathrm{dd}},C_\ell^{\mathrm{ss}}$. However, given our focus on the $H_0 - \Delta M$ degeneracy, these extensions are beyond the scope of the present work. 

Finally, a number of cosmological parameters are kept fixed in our analysis, including $A_s$, $\Omega_b$ and $\Omega_r$, which have significant impact on the shape and amplitude of the power spectrum. Whilst these parameters can also be included in the fit, in this work we assume that one can input them from very precise measurements of the CMB monopole and  anisotropies, and use this information to calibrate low-redshift distance indicators. These modelling choices are equivalent to anchoring the distance indicators to early Universe (well-measured) clustering information rather than to the low-redshift distance ladder, or to BAO distances calibrated using the CMB sound horizon (the standard \textit{inverse distance ladder}). Whilst this approach is not the most general, and in particular assumes that the CMB determination of the clustering parameters is unbiased, we believe it provides a straightforward and important consistency test of the standard cosmological model. In principle, the clustering-based likelihood could instead be combined with low redshift measurements of the power spectrum, thus becoming CMB-independent. Similarly, one could include in the analysis an external prior on $\Delta M$, analogous to the role of the absolute magnitude calibration in the standard distance--redshift relation. Indeed, distance measurements alone are also degenerate in the $H_0$--$M$ plane, and this degeneracy is broken only once a choice of calibrator is specified. A natural extension of this work would therefore be a hybrid analysis in which the clustering information provides an independent geometric constraint, while a prior on $\Delta M$, for example from Cepheid-calibrated SN~Ia, anchors the absolute luminosity scale. Such an approach would be expected to perform at least as well as the standard distance ladder alone, while also providing an additional consistency check on the calibration.


To summarise, we introduced a novel technique that constrains the systematic biases induced by mis-calibration of distance indicators leveraging their clustering properties. For illustrative purposes, we focused on a SN~Ia analogy with a constant $\Delta M$. However, we anticipate that this methodology can be extended to incorporate more general distance indicators and parametrisation of systematic biases. Our forecasts show that the answer to the question \textit{systematic or new physics?}, at least in the the context of the Hubble tension, could potentially be addressed in the foreseeable future.

\section{Acknowledgments}
\begin{acknowledgments}
It is a pleasure to thank Prof. Chris Blake for useful comments and suggestions and for critically reading earlier versions of this manuscript.
This research was conducted in part by the Australian Research Council Centre of Excellence for Gravitational Wave Discovery (project number CE230100016) and funded by the Australian
Government.
\end{acknowledgments}

\appendix

\section{Angular cross-correlations} 
\label{app:Cell_theory}
Following Refs.~\cite{Giani_cross,2018JCAP...10..047S}, we can compress statistical information of a cosmological field into a spherical harmonics-Fourier basis with coefficients:
\begin{equation}
  a_{\ell m}^{i}=4\pi i^\ell\int \frac{d^3k}{\left(2\pi\right)^3} j_{\ell}\left(k \chi(z)\right)\delta_{i}(k,z)Y_{\ell m}^*(\hat{k} )\;,
\end{equation}
where $\chi$ is the FLRW comoving distance and $j_\ell$ are the spherical Bessel function of the first kind. 
The joint two-point statistics of these coefficients, assuming homogeneity and isotropy and expanding in Fourier space, define the angular cross-power spectrum $C_\ell^{AB}$:
\begin{equation}\label{eq:Cell_expl}
\begin{split}
    &\left\langle a_{\ell m}^{A}\;,\; a_{\ell' m'}^{B^*} \right\rangle\equiv        \delta_{\ell\ell'}\,\delta_{mm'}\,C_\ell^{AB}\;\\
     &=\delta_{\ell\ell'}\,\delta_{mm'}\iint d^3\vec{k} \;d^3\vec{k}'\int_{z_A^{\min}}^{z_A^{\max}} dz \int_{z_B^{\min}}^{z_B^{\max}} dz'b_A(z) b_B(z') \\
     &\times \frac{dn_A^{\rm{obs}}(z)}{dz}\frac{dn^{\rm{obs}}_B(z')}{dz'}\langle\delta(k,z),\delta(k',z')\rangle\; j_{\ell}(k\chi_A) j_{\ell'}(k'\chi_B) \;,
\end{split}
\end{equation}
where $k=|\vec{k}|$, and we assumed that the density contrast of a given tracer $\delta_i$ is related to underlying cosmological dark matter distribution  $\delta_m$ via a bias parameter $\delta_i(k,z) = b_i(z) \delta_m(k,z)$.
Since different $k$-modes are independent, $\langle\delta(z,k),\delta(z',k')\rangle=\delta\left(k-k'\right)P_m(k,z,z')$, we can write
\begin{equation}\label{eq:C_ell_pm}
\begin{split}
C_\ell^{AB}&=\int dk \;k^2 \int_{z_A^{\min}}^{z_A^{\max}} dz \int_{z_B^{\min}}^{z_B^{\max}} dz'b_A(z) b_B(z') \\
     &\times \frac{dn_A^{\rm{obs}}(z)}{dz}\frac{dn^{\rm{obs}}_B(z')}{dz'}P_m(k,z,z') j_{\ell}(k\chi_A) j_{\ell'}(k'\chi_B) \;.
\end{split}
\end{equation}    
These expressions can be further simplified by making use of the Limber approximation, which evaluates integrals containing the products of Bessel functions via
\begin{equation}\label{eq:Limber_approx}
\lim_{\epsilon \rightarrow 0} \int dx\;\sqrt{\frac{2x}{\pi}} e^{\epsilon\left(x - \ell\right)} j_{\ell-\frac{1}{2}}(x)f(x)  \approx f(\ell) +\mathcal{O}\left(\ell^{-2}\right)\;,    
\end{equation}
see for example Ref.\cite{LoVerde:2008re} for a formal derivation. Using the above in Eq.~\eqref{eq:C_ell_pm} we obtain
\begin{equation}\label{eq:C_ell_limber}
    C_\ell^{AB}\approx \int_{\Delta z_{\rm{ov}}}dz\frac{ H(z)}{\chi^2(z)}\; b_Ab_B \frac{dn_A^{\rm{obs}}}{dz}\frac{dn^{\rm{obs}}_B}{dz}P_m(k_\ell,z)\;,
\end{equation}
where $k_\ell=(\ell + 1/2)\chi$.

\subsection*{$C_\ell$ in distance and redshift space} 
We want to evaluate the cross-correlation between tracers in different spaces, nominally redshift and luminosity distances. Eqs.~\eqref{eq:C_ell_pm},\eqref{eq:C_ell_limber} can be easily generalised using the Jacobian of the transformation between the two spaces in a given cosmology:
\begin{equation}
   dz = \frac{H(z)}{c}d\chi\;, \qquad d_L = (1+z)\chi\;.
\end{equation}
We assume spectroscopic redshift measurements, so the redshifts uncertainties are negligible compared to the redshift bins width. However, the uncertainty on distance measurements can cause leaking of individual events outside their true distance bin, so that $dn_\mathrm{d}^{\rm{obs}}/d d_L$ and the true distribution $dn_\mathrm{d}^{\rm{true}}/d d_L$ are related by:
\begin{equation}
\frac{dn_{\rm{d}}^{\rm{obs}}}{dd_L}=\frac{dn_{\rm{d}}^{\rm{true}}}{dd_L}W_{\rm{d}}(d_L,\theta,\phi)\; P_i^{d_L}(z,H_0) 
\end{equation}
where $W_{\rm d}$ is the window function that accounts for the survey geometry and completeness, and $P_i^{d_L}$ is the probability that a source at redshift $z$ is assigned to the observed distance bin $d_{L,i}=[d_{L,i}^{\rm obs}, d_{L,i+1}^{\rm obs}]$. Assuming Gaussian distance uncertainties, this bin-assignment probability is
\begin{equation}\label{eq:Pdet}
\begin{split}
P_i^{d_L^{\mathrm{obs}}}(z,H_0)
=
\Phi\left(\frac{d^{\rm obs}_{L, i+1}-d_L(z,H_0)}{\sigma_{d_L}}\right)\\
-
\Phi\left(\frac{d_{L,i}^{\mathrm{obs}}-d_L(z,H_0)}{\sigma_{d_L}}\right).
\end{split}
\end{equation}

\subsection*{Impact of $\Delta M$ on $C_\ell^{\mathrm{g}_i\mathrm{d}_j}$}
The mapping $d_L^{\rm{obs}}=d_L^{\rm{true}}\alpha_{\Delta M}$ affects the cross-correlation $C_\ell^{\rm{g}_i\rm{d}_j}$ by altering the relation between the \textit{true} Fourier modes $k' \rightarrow \alpha k'$, so that $\langle\delta(k),\delta(\alpha k')\rangle = \delta(k-\alpha k')P_m(k)$. As a result, the cross-correlation with the systematically biased distance tracer becomes:
\begin{equation}\label{CellofDeltaM}
\begin{split}
C_\ell^{\mathrm{g}_i \mathrm{d}_j}=4\pi\int_{z_i^{\min}}^{z_i^{\max}} dz \int_{z_j^{\min}(d^{\rm obs})}^{z_j^{\max}(d^{\rm obs})} dz' W_\mathrm{g}^i(z) W_\mathrm{d}^j(z')\\ 
\int_0^\infty dk \; k^2 P_m(k,z,z') j_\ell(k  \chi(z')\alpha_{\Delta M}) j_\ell(k\chi(z))\;,
\end{split}
\end{equation}
which evaluated using the Limber approximation results in Eq.~\eqref{eq:Cell_Limber_DM}.


\section{Density contrast in observed distance space}
\label{app:delta}
The transformation relating the density contrast in observed distance space, in a similar fashion to  Redshift Space Distortion (RSD) analysis (see e.g. Ref.~\cite{1987MNRAS.227....1K}), can be obtained by invoking number conservation of the discrete tracers over the comoving volume:
\begin{equation}\label{eq:N_conservation}
    n_{\rm obs}(d_{\rm obs})\,dV_{\rm obs}
    =
    n_{\rm true}(d_{\rm true})\,dV_{\rm true}.
\end{equation}
To simplify the notation, let us write $\mathbf{x}=\chi_{\rm{true}}$ and $\chi_{\rm{obs}}\equiv \mathbf{s}$, with $\mathbf{s}= \alpha_{\Delta M}\mathbf{x}$.
The Jacobian of the coordinate transformation is
\begin{equation}
    J_{\Delta M}
    \equiv
    \left|
    \frac{\partial \mathbf{s}}
         {\partial \mathbf{x}}
    \right|
    =
    \alpha_{\Delta M}^3 \;,
\end{equation}
which implies that $dV_{\rm obs}=J_{\Delta M}dV_{\rm{true}}$. 
Writing the density fields in terms of their mean densities and density contrasts, $n=\bar{n} \left(1+\delta\right)$, and noticing that mean transforms according to
\begin{equation}
    \bar n_{\rm obs}(\mathbf{s})
    =
    \alpha_{\Delta M}^{-3}
    \bar n_{\rm true}\left(
        \frac{\mathbf{s}}{\alpha_{\Delta M}}
    \right)\;,
\end{equation}
we can finally write
\begin{equation}
    1+\delta_{\rm obs}(\mathbf{s})
    =
    1+
    \delta_{\rm true}\left(
        \frac{\mathbf{s}}{\alpha_{\Delta M}}
    \right),
\end{equation}
from which Eq.~\eqref{eq:delta} follows.
Interestingly, unlike in the RSD case, a constant magnitude calibration systematic does not introduce an additional multiplicative contribution to the density contrast, and only remaps the radial coordinate at which the true density field is evaluated.

\section{Gaussian covariance}
\label{app:Covariance}
For Gaussian fields, the covariance between $C_\ell$'s has a simple analytic form, accounting for  both cosmic variance and finite number of measurements (see for example \citet{Krause_2017}):
\begin{widetext}
\begin{equation}
\begin{split}
\mathrm{Cov}^{\mathrm{G}}\!\left(C^{ij}_{AB}(\ell_1), C^{kl}_{CD}(\ell_2)\right)
&=
\frac{4\pi \delta_{\ell_1 \ell_2}}{\Omega_s (2\ell_1+1)\Delta \ell_1}
\left[
\left(C^{ik}_{AC}(\ell_1)+N_{AC}^{ik}\right)
\left(C^{jl}_{BD}(\ell_2)+N_{BD}^{jl}\right)\right.\\
&\left.\qquad +\left(C^{il}_{AD}(\ell_1)+N_{AD}^{il}\right)
\left(C^{jk}_{BC}(\ell_2)+N_{BC}^{jk}\right)
\right],
\end{split}
\end{equation}
\end{widetext}ƒ
where, to simplify the notation, we used upper indexes $\left\{i,j,k,l\right\}$ for the bin membership and capital lower indexes $\left\{A,B,C,D\right\}$ for different tracers. The noise fields ($\mathcal{N}$) in the expressions above for the number count of independent tracers reduce to their shot noises:
\begin{equation}
\mathcal{N}^{pq}_{XY} \equiv \delta_{pq}\delta_{XY} \frac{1}{\bar{n}^p_X}\;,  \qquad \bar{n}^p_X=\frac{4\pi f_{\rm{sky}}}{N^p_X}\;,
\end{equation}    
above are the shot noises of the fields, which for independent tracers are given by
\begin{equation}
    \mathcal{N}^i_X \equiv \frac{1}{\bar{n}^i_X}=\frac{4\pi f_{\rm{sky}}}{N^i_X}\;,
\end{equation}
where $N^i_X$ is the number of tracers $X$ per steradian in bin $i$. However, since in our construction the tracers $\left\{\rm{s},\rm{d}\right\}$ contain the same objects, their cross-shot noise term is in general non-vanishing in overlapping bins. In particular, we have:
\begin{equation}
    \mathcal{N}^{ij}_{\rm{d}\rm{s}}= \frac{\bar{n}_{\rm{ov}}^{ij}}{\bar{n}^i_{\rm{d}}\bar{n}^j_{\rm{s}}}\;,
\end{equation}
where $\bar{n}_{\rm{ov}}^{ij}$ is the number density of objects in common to both bins $i$ and $j$. Combining the above definitions we obtain Eq.~\eqref{eq:covariance}, in which for simplicity we collapsed pairs of bin-tracers indexes into a single one, e.g. $\left\{X,i\right\}\rightarrow X_i$. 

\section{Fisher information}
Considering the three choices of data-vectors,
\begin{equation}
\mathbf{D}
\in
\left\{
\left\{C_\ell^{\mathrm{s}_i\mathrm{d}_j}\right\},
\left\{C_\ell^{\mathrm{g}_i\mathrm{d}_j}\right\},
\left\{C_\ell^{\mathrm{g}_i\mathrm{d}_j},C_\ell^{\mathrm{g}_i\mathrm{s}_j}\right\}
\right\}
\quad \forall\; \ell,i,j ,
\end{equation}

for a given choice of $\mathbf{D}$, the Fisher matrix is
\begin{align}
\label{eq:fisher}
F_{\alpha\beta}
=
\sum_{\mu,\nu}
\frac{\partial D_\mu}{\partial p_\alpha}
\left[\mathrm{Cov}(\mathbf{D})^{-1}\right]_{\mu\nu}
\frac{\partial D_\nu}{\partial p_\beta},
\end{align}
where $D_\mu$ denotes an element of the flattened data vector, including its multipole and bin indices, and $p_\alpha\in\{H_0,\Omega_m,\Delta M, b_{\mathrm{g}}, b_{\mathrm{d}}\}$. The Fisher matrix provides an estimate of the lower bound on the parameter uncertainties under the Gaussian approximation. For a multi-parameter analysis, the marginalised uncertainty forecast for parameter \(p_\alpha\) is
\begin{align}
    \sigma(p_\alpha) = \sqrt{\left(F^{-1}\right)_{\alpha\alpha}}\, ,
\end{align}
which corresponds to the Cramér--Rao lower bound \citep{rao1945information,cramer1999mathematical}.

Our final forecasts are quoted in the $\{H_0,\Delta M\}$ subspace by taking the corresponding $2\times2$ block of $F^{-1}$, which effectively marginalises over $\{\Omega_m, b_{\mathrm{g}}, b_{\mathrm{d}}\}$.  In the Fisher--simulation comparison, we instead fix these nuisance parameters and vary only $H_0$ and $\Delta M$.

\bibliography{apssamp}

\end{document}